\documentclass[onecolumn,showpacs,prd,aps,amsmath,nofootinbib,amssymb,eqsecnum,superscriptaddress]{revtex4-1}
\usepackage{mathrsfs} % -> \mathscr

\usepackage{color}
\usepackage{hyperref}
\usepackage{graphicx}
\usepackage{bm} % bold math
\usepackage{mathrsfs} % -> \mathscr
\allowdisplaybreaks
\newcommand{\be}{\begin{equation}}
\newcommand{\ee}{\end{equation}}
\newcommand{\bea}{\begin{eqnarray}}
\newcommand{\eea}{\end{eqnarray}}

\newcommand{\Msun}{M_{\odot}}

\newcommand{\nn}{\nonumber}
\newcommand{\pp}{\varphi}
\newcommand{\dd}{\mathrm{d}}

\def\spose#1{\hbox to 0pt{#1\hss}}
\def\lta{\mathrel{\spose{\lower 3pt\hbox{$\mathchar"218$}}
          \raise 2.0pt\hbox{$\mathchar"13C$}}}
\def\gta{\mathrel{\spose{\lower 3pt\hbox{$\mathchar"218$}}
          \raise 2.0pt\hbox{$\mathchar"13E$}}}

\begin{document}

\title[Cosmic background radiation in the vicinity of a Schwarzschild black hole: no classic firewall]{Cosmic background radiation in the vicinity of a Schwarzschild black hole:\\ no classic firewall}

\date{03 December 2014} 

\author{M. Wielgus}\email[]{maciek.wielgus@gmail.com}
 \affiliation{Nicolaus Copernicus Astronomical Center, ul. Bartycka 18, PL-00-716
               Warszawa, Poland}      
               \affiliation{Institute of Micromechanics and Photonics, Warsaw University of Technology, ul. \'{S}w. A. Boboli 8, PL-02-525
               Warszawa, Poland}         
\affiliation{Institute of Physics, Silesian University in Opava,
                   Bezru{\v c}ovo n{\'a}m. 13, CZ-746-01 Opava,
                   Czech Republic}                             
\author{ G. F. R. Ellis}
\affiliation{ Mathematics Department, University of Cape Town,
    Rondebosch, Cape Town 7701, South Africa}
\author{F. H. Vincent}
\affiliation{Nicolaus Copernicus Astronomical Center, ul. Bartycka 18, PL-00-716
               Warszawa, Poland}
               \author{M. A. Abramowicz}
\affiliation{Institute of Physics, Silesian University in Opava,  Bezru{\v c}ovo n{\'a}m. 13, CZ-746-01 Opava,
                   Czech Republic}                             
\affiliation{Nicolaus Copernicus Astronomical Center, ul. Bartycka 18, PL-00-716
               Warszawa, Poland}
\affiliation{Department of Physics, University of Gothenburg, SE-412-96 G{\"o}teborg, Sweden}

\begin{abstract}
The Cosmic Blackbody Background Radiation
pervades the entire Universe, and so falls into every astrophysical black hole. The blueshift of the infalling photons, measured
by a static observer, is infinite at the event horizon. This raises a question as to whether a ``firewall'' of high energy density
may form just outside the horizon, or whether the effect can be attributed exclusively to a singular behavior of the
static observer's frame at the horizon. In principle, the presence of such firewall may alter the motion of the infalling
matter, influence the black hole evolution, or even invalidate the {\it vacuum} Einstein field equation solution as a realistic
approximation for black holes.   In this paper we show by means of analytic calculations that all these effects indeed exist,
but their magnitude is typically negligibly small, even though the matter stress tensor is divergent in the static frame at $r=2M$.
That is not surprising because of the divergent relation of that frame to a freely falling frame as $r \rightarrow 2M$; however it represents a kind of  classical analogue for the Black Hole Complementarity principle that has been proposed for quantum effects near a black hole. What is perhaps more surprising is the divergence of the radiation stress tensor for massive particles moving on circular geodesic orbits for values of $r$ approaching $r = 3M$. However such orbits will not occur for infalling matter in realistic accretion discs.
\end{abstract}


\maketitle
\section{Introduction}

\label{section-Introduction}

In this paper, we investigate several aspects of the interplay between the Schwarz\-schild black hole and the Cosmic Blackbody Radiation (CBR). The Universe is filled with the CBR, which was emitted from the Hot Big Bang era. Its present temperature is \citep{WMAP},
%--------------------------------------------------------------
\begin{equation}
T_{\rm CBR} = 2.73\,[{}^\circ {\rm K}].
\label{CBR-temperature}
\end{equation}
%--------------------------------------------------------------
The present Hubble time, i.e., the ``age of the Universe'' is \citep{WMAP},
%--------------------------------------------------------------
\begin{equation}
t_{\rm Hubble}  = 1.37 \times 10^{10}\,[{\rm yrs}].
\label{Hubble-time}
\end{equation}
%--------------------------------------------------------------
The ``astrophysical'' black holes, i.e., these  about which astrophysicists have observational data, belong to three classes\footnote{The existence of the intermediate-mass black holes is neither firmly established, nor commonly accepted. The majority opinion is against its reality. The brilliant recent estimate $M = 400 M_{\odot}$ for a black hole in the M82 cluster comes not from the unquestionable Kepler-law-based {\it measurements}  (as in the stellar and supermassive cases), but from a far less certain argument based on scaling properties of the so-called 3:2 twin peak QPOs (see \citep{Nature32, Abramowicz32}). Despite long lasting and continous efforts, no observational indications for ``primordial'' mini black holes, with $M \ll M_{\odot}$ have been found (see e.g. \citep{Abramowicz2009, Carr2010}).}, depending on their {\it observed} masses $M$,
%--------------------------------------------------------------
\begin{subequations}
%\label{equations}
\begin{align}
{\rm stellar} && M \sim 10 M_\odot \label{astrophysical-BH-stellar} \\
{\rm intermediate} && M \sim \{10^2 M_\odot - 10^{4} M_\odot \}  \label{astrophysical-BH-inter} \\
{\rm supermassive} && M \sim \{10^6 M_\odot - 10^{10} M_\odot \}.
\label{astrophysical-BH-super}
\end{align}
\end{subequations}
%--------------------------------------------------------------
Here $M_\odot = 1.99 \times 10^{33}\,[{\rm g}]$ is the mass of the Sun.\\
The black  hole mass $M$ determines several characteristic scales relevant for our discussion. In particular, for the Schwarzschild black hole it is,
%--------------------------------------------------------------
\begin{subequations}
\label{equations}
\begin{align}
{\rm size}\,R_{\rm G}&=& 
\frac{2GM}{c^2} = 2.95 \times 10^5 \left( \frac{M}{M_\odot}\right)\, {\rm [cm]}
\label{BH-size}                         
\\
{\rm area}\,A_{\rm G}&=&
4\pi \left(R_{\rm G} \right)^2= 1.10 \times 10^{12} \left( \frac{M}{M_\odot}\right)^2\, {\rm [cm]}^2 
\label{BH-area}     
\\          
{\rm time}\,t_{\rm G}&=& 
4\pi \frac{R_{\rm G}}{c} = 1.23 \times 10^{-4} \left( \frac{M}{M_\odot}\right)\, {\rm [s]}
\label{BH-time}     
\\          
{\rm Hawking~temperature}\,T_{\rm H}&=& 
\frac{\hbar c^3}{8\pi G k M} = 6.17 \times 10^{-8} \left( \frac{M}{M_\odot}\right)^{-1}\, {\rm [}^\circ\rm{K]} 
\label{H-temperature}    
\\          
{\rm Hawking~power}\,L_{\rm H} &=& 
(\sigma T_{\rm H} ^4) A_{\rm G}
  = 9.02 \times 10^{-22} \left( \frac{M}{M_\odot}\right)^{-2}\, {\rm [erg/sec]} 
\label{H-power}    
\\          
{\rm Hawking~time}\,t_{\rm H}&=& 
\frac{Mc^2}{L_{\rm H}} = 6.28 \times 10^{67} \left( \frac{M}{M_\odot}\right)^{3}\, {\rm [yrs]}
\label{H-time}    
\\          
{\rm Eddington~power}\,L_{\rm E}&=& 
\frac{4\pi G m_p c M}{\sigma_T} = 1.26\times 10^{38} \left( \frac{M}{M_\odot}\right)\, {\rm [erg/s]} 
\label{E-power} 
\\  
{\rm Eddington~temperature}\,T_{\rm E}&=& 
\left(\frac{L_{\rm E}}{ A_{\rm G} \sigma}\right)^{1/4} = 3.77\times 10^{7} \left( \frac{M}{M_\odot}\right)^{-1/4}\, {\rm [erg/s]} 
\label{E-temperature} 
\\  
{\rm Eddington~time}\,t_{\rm E}&=& 
\frac{M c^2}{L_{\rm E}} = 4.51 \times 10^{8}\, {\rm [yrs]}  < t_{\rm Hubble}
\label{E-time} 
\\          
{\rm CBR~power}\,L_{\rm CBR}&=& 
(\sigma T^4_{\rm CBR})A_{\rm G} = 3.45 \times 10^9 \left( \frac{M}{M_\odot}\right)^{2}\, {\rm [erg/s]} 
\label{CBR-power} 
\\  
{\rm CBR~time}\,t_{\rm CBR}&=& 
\frac{M c^2}{L_{\rm CBR}} = 1.64 \times 10^{37} \left( \frac{M}{M_\odot}\right)^{-1}\, {\rm [yrs]}  
\label{CBR-time} 
\end{align}
\end{subequations}
%--------------------------------------------------------------
Symbols in (\ref{BH-size})-(\ref{CBR-time}) have their standard meaning, e.g. $G =$ Newton's gravitational constant, $\hbar = $ Planck's constant, $\sigma =$ Stefan-Boltzman's constant, $\sigma_T$ Thomson scattering cross section. The Eddington luminosity $L_{\rm E}$ is a~convenient scale for accretion radiative power --- accretion disks have their luminosities of the order or smaller than $L_{\rm E}$, \citep{Abramowicz2005}. The Edington time $t_{\rm E}$ estimates the timescale of a black hole evolution due to accretion of matter: in the time $t_{\rm E}$, a black hole will (roughly) double its mass due to accretion.

From equations (\ref{CBR-temperature})-(\ref{Hubble-time}), (\ref{H-temperature})-(\ref{H-time}) and (\ref{E-time})  a well known conclusion yields --- that for the astrophysical black holes it is,
%--------------------------------------------------------------
\begin{equation}
\boxed{
~T_{\rm H} \ll T_{\rm CBR} ,~~t_{\rm H} \gg t_{\rm Hubble}  ~~~{\rm and}~~~t_{\rm H} \gg t_{\rm E}~
}
\label{Hawking-inequalities}
\end{equation}
%--------------------------------------------------------------
i.e. that the astrophysical black holes are much (orders of magnitude) cooler than the thermal CBR bath in
which they are immersed, and therefore they should radiate no Hawking radiation. Even if they would, the time-scale of their evaporation would be absurdly long --- orders of magnitude longer than the Hubble time. Thus, assuming correctness of the standard Einstein's general relativity, and of the original semi-classical Hawking's arguments which lead to (\ref{H-temperature}), one may be tempted to conclude that {\it Hawking's radiation plays no role for the astrophysical black holes.}

However, the issue here is more subtle. Hawking radiation introduces {\it a matter of principle problem} of a~fundamental importance for physics: the information paradox --- with Hawking's radiation and black hole evaporation, the black hole evolution cannot be unitary \citep{Hawking1976, Hawking2005}. Considerable excitement has followed a recent suggestion that ``the only solution'' of the paradox  may be given by a Planck-scale (size, density) ``firewall'' that should form, as it was claimed, at the black hole horizon due to both (a) infinite blueshift of the Hawking radiation photons and (b) their quantum entanglement\footnote{Almheiri et al. \citep{Firewall} argue that two standard assumptions made in discussions of quantum properties of black holes, namely, that “(i) {\it Hawking radiation is in a pure state}, (ii) {\it the information carried by the radiation is emitted near the horizon, with low energy effective field theory valid beyond some distance from the horizon}, are incompatible with a statement that (iii) {\it the infalling observer encounters nothing unusual at the horizon}. Here we pick up and stress just one crucial issue in their arguments --- the ``infinite blueshift'' at the horizon. Other authors, e.g., \cite{Firewalllimit}, picked up other problems.}. It was also claimed \citep{Firewall,Firewall1} that the firewall would burn up any in-falling object at the horizon. 

We will not discuss the issue of the quantum firewalls here. Instead, we ask the question whether somehow similar ``classical''' firewalls could exist.  This question arises from the very obvious remark that the crucial ingredient of the quantum firewall arguments is the ``infinite blueshift'' of the Hawking radiation at the black hole event horizon. Such infinite blueshift is measured by the ``zero angular momentum observers'' (ZAMO), who are static observers in the Schwarzschild spacetime\footnote{The ZAMO observers are accelerated; they do not follow geodesic lines. They are mathematically convenient, as they naturally (but not uniquely) foliate the Schwarzschild and Kerr space-times, thus providing definitions of ``space'', ``time'' and ``rest frame''. In addition, in the case of static space-times (e.g. Schwarzschild), they embrace the Killing time symmetry, as their trajectories coincide with the trajectories of the time-like Killing vector.  Thus they define a~geometrically preferred rest frame. Further we refer to ZAMO simply as ``static observer'', since we limit these investigations to the static Schwarzschild spacetime.}. Obviously, the infinite blueshift in the static observer's frame occurs at the black hole event horizon for {\it all} photons, also these which originate from standard and familiar astrophysical situations --- the infinite bluesift is not peculiar to the quantum entanglement of Hawking radiation. Would the infinite blueshift of these ``clasical'' photons form a ``classical'' firewall at the black hole horizon? 

Here, as an example, we will only consider the CBR photons. Not only because of their fundamental importance, but also because of their well-known properties and of unquestionable presence. While characteristics of photons which originate due to local accretion depend primarly on specific, unknown, local circumstances related to specific sources, the entire Universe is pervaded the CBR with {\it explicitly known} properties. At the event horizon, the CBR photons experience infinite blueshift in the static observer's frame. Consequently the energy density of the radiation as measured by such observers diverges as $r \rightarrow 2M$. A classic firewall (shell of extremely energetic photons) is formed in the static observer's frame. What are the astrophysical consequences of this?  In particular, what do freely in-falling objects experience as they cross this firewall? Is there any back-reaction effect on the black hole itself due to this hypothetical firewall of infinitely blue-shifted CBR in-falling radiation? 

One should be aware of the fact that the question about a possible importance of the CBR for the black hole evolution due to accretion is {\bf NOT} directly relevant for the question of the ``burning the infalling objects'' aspect of the firewall physics. Indeed, one may easily conclude from equations (\ref{E-temperature}), (\ref{CBR-temperature}) and  (\ref{E-time}), (\ref{CBR-time}) that effect of the CBR accretion is by many orders of magnitude smaller than the ordinary Eddington accretion,
%--------------------------------------------------------------
\begin{equation}
\boxed{
~T_{\rm CBR} \ll T_{\rm E} ,~~t_{\rm CBR} \gg t_{\rm E},  ~~~({\rm however}~~~t_{\rm CBR} \ll t_{\rm H})~
}
\label{CBR-inequalities}
\end{equation}
%--------------------------------------------------------------
Exactly like the ``Hawking inequalities'' (\ref{Hawking-inequalities}) do not imply whether the (hypothetical) Hawking firewall would burn the infalling objects, the similar ``CBR inequalities'' (\ref{CBR-inequalities})  do not imply whether the (hypothetical) CBR firewall would burn (or slow down) the infalling objects. In both cases a more detailed analysis is needed.

Motivated by this, we first performed analytic calculations of the CBR stress-energy tensor at an arbitrary distance from the event horizon (located at $r =R_{\rm G}$ in the Schwarzschild coordinates), showing that not only the CBR temperature diverges, as expected, in the static observer's frame as $r \rightarrow 2M$, but also it's stress energy tensor in that frame diverges their. Putting this divergent stress tensor in the field equations might perhaps lead to a spacetime singularity; however we show this is not the case.   We investigated the influence of the CBR on the dynamics of material particles in the black hole vicinity as they fall in, characterizing the classic firewall and the energy that it can deposit on an in-falling material object. We discuss the back-reaction of the radiation stress-energy tensor on the metric, showing how it is negligible on small timescales. Additionally, we calculate the total rate of mass increase due to the in-falling radiation and find the non-stationary Vaidya spacetime that partly accounts for the stress-energy tensor of the CBR field. Our conclusion is that the classical CBR firewall ``exists'' in principle but has no significant effect on freely infalling observers or on the evolution of the black hole. This is perhaps not surprising in that it relates to the divergent behaviour of the ZAMO rest frame as $r \rightarrow 2M$. However our analysis also shows that for radial observers moving at $r =r_C = constant$, the CBR energy momentum diverges as $r_c\rightarrow 3M$: this happens quite outside the horizon at $r=2M$. This divergence therefore cannot be associated with the singular properties of the ZAMO frame as $r \rightarrow 2M$. However infalling particles will not `naturally' move on these geodesics. Thus this divergence also will not significantly affect real black hole dynamics.    

Black hole firewalls of any kind are not possible without the ``infinite blueshift'' of photons as measured at the black hole horizon by the static observers. In this paper we have proven that, according to the standard Einstein general relativity theory, no classic CBR firewalls will be formed. Possible formation of Hawking radiation firewalls is still under debate, despite arguments against it mentioned earlier. Arguments presented here should convince the firewall enthusiasts that the ``infinite redshift'' is only a necessary, but {\bf NOT} a sufficient condition for all postulated firewalls, with or without the quantum entanglement.

% % % % % % % % % % % % % % % % % % % % % % % % % % % % % % % % % % % % % % % % % % % % % % % % % % % % % % % % % % % % %
\section{CBR stress-energy tensor}\label{sec:Tab}
% % % % % % % % % % % % % % % % % % % % % % % % % % % % % % % % % % % % % % % % % % % % % % % % % % % % % % % % % % % % % %
Let us consider a sphere of %infinite
radius $r_0 \gg 2M$ in a~Schwarzschild spacetime
that emits radiation isotropically in its rest frame. We will now compute the stress-energy tensor $T^{\mu\nu}$ of such a~radiation field at any point outside of the event horizon. This is a very symmetric problem, closely resembling one considered by \citet{abramowicz90}, i.e., uniformly radiating static spherical object in the Schwarzschild spacetime. While some results can be deduced based on the findings of \citep{abramowicz90}, we choose a~systematic self-contained approach to our calculations, performing them from first principles.  \\

The Schwarzschild metric is given by
\be
\dd s^2 = g_{tt} \dd t^2 + g_{rr} \dd r^2 + r^{2}\left(\dd \theta^2 + \mathrm{sin}^2\theta \dd\varphi^2\right) \ \text{,}
\ee
where the signature is $(-,+,+,+)$, 
\begin{equation}\label{eq:Schw1}
g_{tt} = -(1 -2M/r),\,\, g_{rr} = -g_{tt}^{-1}
\end{equation} and $G=c=1$. The Schwarzschild basis is
$(\mathbf{\partial}_t,\mathbf{\partial}_r,\mathbf{\partial}_\theta,\mathbf{\partial}_\pp)$
where $\boldsymbol{\partial}_t$ and $\boldsymbol{\partial}_\pp$ are Killing vectors associated with temporal and azimuthal symmetries.  The orthonormal basis of the static observer is $(\mathbf{u}_S,\mathbf{e}_r,\mathbf{e}_\theta,\mathbf{e}_\pp)$ where $\mathbf{e}_i = \mathbf{\partial}_i / \sqrt{g_{ii}}$ and \begin{equation}\label{eq:static}
\mathbf{u}_S = (\sqrt{-1/g_{tt}},0,0,0)
\end{equation} is the static observer's four-velocity.

\subsection{Intensity}
Due to spacetime symmetries, the following quantities are conserved along
a null geodesic with tangent four-vector $\boldsymbol{u}$:
\bea
E &=& -u_t, %\\ \nn
L = u_\pp \ \text{.} %\\ \nn
\eea
The motion of photons in Schwarzschild spacetime is planar because of the spherical symmetry. Assuming
without loss of generality $u^{\theta}=0$ and $\theta=\pi/2$, the equation of motion reads
\be
\frac{1}{L^2}\left(\frac{\dd r}{\dd \lambda}\right)^2 + U(r) =l^{-2} \ \text{,}
\ee
where $\lambda$ is an affine parameter, $ l=L/E$ is the rescaled
angular momentum and
\be
U(r)=\frac{1-2M/r}{r^2} \ \text{.}
\ee
The potential $U(r)$ reaches a~maximum $U_M=1/27$ at $r=3M$. Let us consider a~photon at some radius $r_0 > 3M $ with $\dd r / \dd \lambda <0$.
It can have any value of $l$ provided $l^{-2}\ge U(r_0)$, thus
$l \in [0,1/\sqrt{U(r_0)}]$. The character of the photon's trajectory depends on the value of $l$, i. e.,
\begin{itemize}
\item if $l^{-2} > U_M$ then the photon will fall into the black hole;
\item if $l^{-2} = U_M$ then the photon can circularize at $r=3M$, but the equilibrium is unstable and any perturbation will send it either to infinity or into the black hole;
 \item if $l^{-2} <U_M$, the photon will go to smaller radii until it reaches a radius $r_m$ defined by $U(r_m)=l^{-2}$. At $r_m$,  $\dd r / \dd \lambda =0$. As it cannot go to smaller $r$ (it hits the potential barrier) it will return to bigger radii and escape to infinity with $\dd r / \dd \lambda >0$.
\end{itemize}

The bolometric intensity $I(r)$ transported along a null geodesic and
measured by a static observer satisfies

\be
\frac{I(r)}{\left(\boldsymbol{u} \cdot \boldsymbol{u}_S\right)^4(r)} = \mathrm{const} \ \text{,}
\ee
where $\boldsymbol{u}$ is the photon  trajectory tangent vector. Thus the intensity measured by an observer at coordinate radius $r$, coming from any direction, is
\be
\label{eq:intens}
I(r) = \frac{I_{\infty}}{g_{tt}^2(r)} \ \text{,}
\ee
where $I_{\infty}$ is the intensity emitted on the sphere at infinity, and as thermal radiation obeys Stefan-Boltzmann law
\be
I_\infty = \sigma T_{CBR}^4 \ \text{,}
\ee
where $\sigma$ denotes Stefan-Boltzmann constant. The cosmic background radiation temperature is presently equal to $T_{CBR} = 2.726 $ K, but was as large as 3000 K at the moment of its emission, which corresponds to an intensity change of 12 orders of magnitude. The current intensity of the CBR is
\be
I_\infty = 3.131\cdot 10^{-3} \ \text{[erg/cm$^2$/s] \ .}
\ee
The expression given in Eq. (\ref{eq:intens}) is valid no matter whether the photon went "straight" to the observer, or orbited many times around the black hole. This is because the blueshift of the infalling radiation depends only on the potential difference between the point of emission and the point of observation. The observer location and photon direction seen at this location are \emph{uniquely} related to the emission  location and direction on the "infinite CBR sphere" (they are connected by a unique null geodesic). As the "infinite CBR sphere" is radiating homogeneously and isotropically with intensity $I_{\infty}$, the observer's sky is also uniformly bright, with the value of the intensity given by Eq. (\ref{eq:intens}), except for a~dark circle due to the presence of the black hole (no photons arrive from those directions). One needs to relate the angular size of the dark sky region to the static observer's coordinate radius $r$.

\subsection{Dark sky region}

Let us label a photon reaching the observer by the angle $a \in [0,\pi]$ between
the $\boldsymbol{e}_r$ vector (pointing outwards) and the incoming
photon tangent vector (this is the same notation as in \citep{abramowicz90}, Fig. 3). This angle is defined so that $a=0$ means a photon moving radially away from the black hole, and $a=\pi$ a photon falling radially
towards the black hole. Thus \textit{$a=\pi$ is the observer's zenith}.
Clearly, there will be some angle $a_0(r)$   such that the sky will be uniformly bright for $a\in[a_0,\pi]$ 
and dark otherwise. Thus $2a_0(r)$ is the  perceived angular diameter of the black hole on the observer's local sky. Our goal is now to derive $a_0(r)$. \\

\begin{figure}[htbp!]
\begin{center}
\includegraphics[trim = 0mm 0mm 0mm 0mm, height=5.9cm]{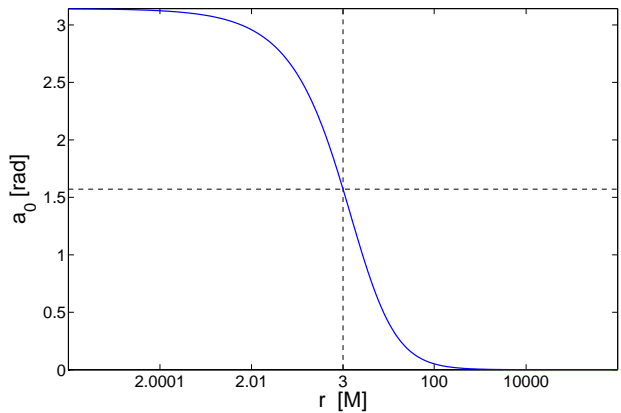}
\caption{The angle $a_0$ as a function of radius. Dashed lines denote the location of the photon orbit, where the dark sky region extends to half of the local sky. }
\label{fig:alpha}
\end{center}
\end{figure}

Let us consider a photon reaching the observer with
angle $a$ and with some angular momentum $l$.
Let us consider the four-vector $\boldsymbol{p}$ equal
to the projection of the photon tangential four-vector $\boldsymbol{u}$ orthogonal
to the static observer four-velocity. It is easy to show that
\be
\boldsymbol{p} = (0,u^i) \ \text{.}
\ee
Moreover
\bea
\boldsymbol{p} &=& \cos a \, \boldsymbol{e}_r + \sin a \, \boldsymbol{e}_\pp = \frac{ \cos a}{\sqrt{g_{rr}}} \boldsymbol{\partial}_r + \frac{ \cos a}{\sqrt{g_{\pp\pp}}} \boldsymbol{\partial}_\pp \\ \nn
&=& u^r \boldsymbol{\partial}_r + u^\pp \boldsymbol{\partial}_\pp = \frac{\dd r}{\dd \lambda} \boldsymbol{\partial}_r + g^{\pp\pp} L \boldsymbol{\partial}_\pp \ \text{.} \\ \nn
\eea
Thus
\bea
\label{eq:cosin}
\cos a &=& \sqrt{g_{rr}} \frac{\dd r}{\dd \lambda}=\pm\sqrt{g_{rr}} L \sqrt{l^{-2}-U} \ \text{,} \\ \nn
\sin a &=&\frac{L}{\sqrt{g_{\pp\pp}}} \ \text{,} \\ \nn
\tan a &=&\pm\sqrt{\frac{l^2 U(r)}{1-l^2 U(r)}}  \ \text{,} \\ \nn
\eea
where the $\pm$ sign is given by the sign of $\dd r / \dd \lambda$ at $r$. If $r \geq~3M$, photons coming to the observer with $\dd r / \dd \lambda<~0$
come from infinity and correspond to bright regions of the sky. Thus all
$a\in[\pi/2,\pi]$ are bright. Photons with $\dd r / \dd\lambda~>~0$ can reach
the observer provided they satisfy $l^{-2}<U_M$. Thus the limiting $a_0$
value is given by
\be
a_0(r \geq 3M) = \arctan \left(\sqrt{\frac{U(r)/U_M}{1-U(r)/U_M}}\right) \in [0,\pi/2] \ \text{.}
\label{eq:a0above3}
\ee
If $r<3M$, no photons can reach the observer with $\dd r / \dd \lambda>~0$,
thus all
$a\in[0,\pi/2]$ are dark. Photons with $\dd r / \dd \lambda<~0$ can reach
the observer provided they satisfy $l^{-2}>~U_M$. Thus the limiting $a_0$
value is given by
\be
a_0(r<3M) = \pi- \arctan \left(\sqrt{\frac{U(r)/U_M}{1-U(r)/U_M} }\right) \in [\pi/2,\pi] \ \text{.}
\label{eq:a0below3}
\ee
For both cases, the sky is bright for
\be
a_{\mathrm{bright}} \in [a_0,\pi] \ \text{.}
\ee
Let us quickly investigate extreme cases. If $r=\infty$ then the 
sky is totally bright except in the direction $a=0$. For $r=3M$, the sky
is bright for $a\in[\pi/2,\pi]$,  thus half-bright. For $r=2M$,
sky is only bright in the direction $a=\pi$ (local zenith) and dark for
all other directions, see Fig. \ref{fig:alpha}.

\subsection{Explicit tensor components}
Knowing the specific intensity at every radius $r >2M$, we can calculate the stress-energy tensor components in the static observer's frame, by integrating the intensity over the observer's local sky. Hereafter we denote a static observer's tetrad components with indices in brackets. We have
\be
T^{(\mu)(\nu)} = \int I(r) \, n^{(\mu)} n^{(\nu)} \dd\Omega \ \text{,}
\label{eq:tmunu}
\ee
where $\boldsymbol{n}$ is a unit spacelike vector obeying
\be
n^{(\mu)} = p^{(\mu)} / p^{(t)}
\ee
and $\dd\Omega$ is the solid angle element. Noting there is only a~contribution from the bright region of the sky, where the  intensity does not depend on the direction, one finds all non-zero components of $T^{(\mu)(\nu)}$ by  elementary integration (cf. \citealt{abramowicz90}, Eqs. 3.31-3.36),
\be
T^{(t)(t)} = 2 \pi I(r) (1 + \cos a_0 ) \ \text{,}
\label{eq:Ttt}
\ee
\be
T^{(t)(r)} = - \pi I(r) \sin^2 a_0  \ \text{,}
\label{eq:Ttr}
\ee
\be
T^{(r)(r)} = \frac{2}{3} \pi I(r) (1 + \cos^3 a_0 )  \ \text{,}
\label{eq:Trr}
\ee
\be
T^{(\theta)(\theta)} = T^{(\phi)(\phi)} = \frac{1}{3} \pi I(r) (2 + 3\cos a_0 - \cos^3 a_0 )  \ \text{.}
\label{eq:Tthth}
\ee
Based on Eqs. (\ref{eq:a0above3})-(\ref{eq:a0below3}) explicit formula for $a_0(r)$ can be given
\be
a_0(r) = \left\{
  \begin{array}{l l}
     \arcsin\left[\frac{3\sqrt{3}M(1-2M/r)^{1/2}}{r} \right] & \quad \text{for $r \geq 3M$}\\
   \pi - \arcsin\left[\frac{3\sqrt{3}M(1-2M/r)^{1/2}}{r} \right]& \quad \text{for $r < 3M$ }
  \end{array} \right.
\ee
It is easy to check that the trace of the $T^{(\mu) (\nu)}$ tensor is zero,
\begin{equation}\label{eqn:trace}
T^{(\mu)}_{\ \ \ (\mu)} = 0 \ \text{,}
\end{equation}
as is expected from the radiation field stress-energy tensor. One can also see that the form of the flux component in Eq. (\ref{eq:Ttr}) corresponds simply to $\nabla_\mu ( T^{\mu \nu} \eta_\nu) = 0$  (which follows because $\eta_\nu$ is the time-symmetry Killing vector).\\

\begin{figure}[htbp!]
\begin{center}
\includegraphics[trim = 5mm 0mm 0mm 0mm, height=5.9cm]{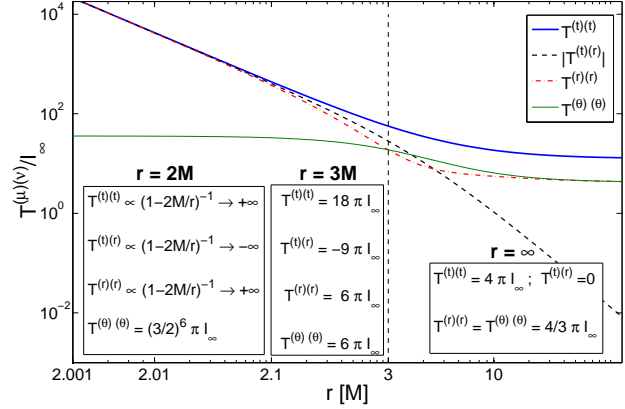}
\caption{Radial dependance of the $T^{(\mu)(\nu)}$ components.}
\label{fig:Tmunu}
\end{center}
\end{figure}

The behavior of the $T^{(\mu) (\nu)}$ tensor in the close vicinity of the horizon is of particular interest to us. The asymptotic behavior of the components as $r \rightarrow 2M$ can be found by utilizing Eqs. (\ref{eq:Ttt})-(\ref{eq:Tthth}),  evaluating the following quantities in the limit of $r \rightarrow 2M$: 
\be
T^{(t)(t)}|g_{tt}| \rightarrow \frac{27}{4} \pi I_\infty  \ \text{,}
\label{eq:TttH}
\ee
\be
T^{(t)(r)}|g_{tt}| \rightarrow - \frac{27}{4} \pi I_\infty \ \text{,}
\label{eq:TtrH}
\ee
\be
T^{(r)(r)}|g_{tt}| \rightarrow \frac{27}{4} \pi I_\infty  \ \text{,}
\label{eq:TrrH}
\ee
\be
T^{(\theta)(\theta)} = T^{(\phi)(\phi)} \rightarrow \left(\frac{3}{2}\right)^6 \pi I_\infty \ \text{.}
\label{eq:TththH}
\ee
Hence $T^{(t)(t)}$, $T^{(t)(r)}$ and $T^{(r)(r)}$ are divergent at the horizon. Note that the asymptotic relation between the tensor components at the horizon corresponds to a point-source stress-energy tensor, for which $T^{(t)(t)} = T^{(r)(r)} = |T^{(t)(r)}|$. Since the bright sky viewing angle $(\pi - a_0)$ goes to zero at the horizon, this should not be surprising. The radial dependence of the $T^{(\mu)(\nu)}$ tensor components, calculated %according to %general formulas 
from 
Eqs. (\ref{eq:Ttt})-(\ref{eq:Tthth}), is illustrated in Fig. \ref{fig:Tmunu}.
Schwarzschild coordinate components $T^{\mu \nu}$ can be calculated with a~simple coordinate substitution.\\

In summary, two opposite effects are present as we approach the event horizon. On the one hand the bolometric intensity diverges with $g_{tt}^{-2}$. On the other hand, the viewing angle of the bright sky region approaches zero. We found that the intensity divergence effect dominates and that photon density $T^{(t)(t)}$, photon flux $T^{(t)(r)}$ and pressure component $T^{(r)(r)}$ diverge in the static observer's orthonormal frame as $r$ approaches $2M$. While the single photon infinite blueshift is a well recognized black hole feature, our calculations indicate the existence of a radiation field that is divergent  in terms of its  energy density.

\subsection{Stress-energy tensor in Eddington-Finkelstein coordinates}
The line element of the Schwarzschild spacetime in Eddington-Finkelstein coordinates takes the following form
\be
\dd s^2 = -\left[1 - \frac{2M}{r} \right] \dd v^2 + 2 \dd v \dd r + r^2(\dd \theta^2 + \sin^2 \theta \dd \phi^2) \ \text{,}
\label{eq:lineEF}
\ee
with advanced null coordinate
\be
v = t -r - 2M \ln(r/2M - 1) \ \text{.}
\ee
Unlike Schwarzschild coordinates, Eddington-Finkelstein coordinates penetrate the black hole horizon and therefore are not prone to the coordinate singularity effects at $r = 2M$. The CBR stress-energy tensor components can be found explicitly,
\be
T^{vv} = \frac{2}{3}\pi I_\infty \left[ \frac{1 + \cos a_0}{1-2M/r} \right]^3 \ \text{,}
\ee
\be
T^{vr} = \frac{1}{3}\pi I_\infty (2 \cos a_0 -1)\left[ \frac{1 + \cos a_0}{1-2M/r} \right]^2 \ \text{,}
\ee
\be
T^{rr} = \frac{2}{3}\pi I_\infty (1 - \cos a_0 + \cos^2 a_0)\left[ \frac{1 + \cos a_0}{1-2M/r} \right] \ \text{.}
\ee
Other components of the $T^{\mu \nu}$ tensor are identical to their counterparts in the Schwarzschild coordinates. Since the quantity in square brackets is finite at the horizon,
\be
\label{eq:BracketFactor}
\left[ \frac{1 + \cos a_0}{1-2M/r} \right] \rightarrow \left(\frac{3}{2} \right)^3
\ \text{,}
\ee
the CBR stress-energy tensor is finite in the Eddington-Finkelstein coordinates.\\

 At this point we may definitely conclude, that the diverging energy density implied by the Eqs. (\ref{eq:Ttt})-(\ref{eq:Trr}) is an effect of the Schwarzschild coordinate singularity at $r = 2M$ where the timelike Killing vector $\boldsymbol{\partial}_t$, which determines the four-velocity of the static observers,  becomes asymptotically null. \\

The remaining problem is to evaluate the magnitude of the CBR-related effects in the vicinity of the horizon. Since black holes do attract CBR photons and influence their trajectories, some growth of the energy density in the vicinity of the horizon should be expected even for observers more physically meaningful than the static one.

% % % % % % % % % % % % % % % % % % % % % % % % % % % % % % % % % % % % % % % % %
\section{Radiation influence on the observers}\label{sec:firewall}
% % % % % % % % % % % % % % % % % % % % % % % % % % % % % % % % % % % % % % % % %

While the infinite value of the radiation energy density at the horizon is attributed to the coordinate singularity, the Schwarzschild coordinates are non-singular for $r = 2M + \epsilon$ for every $\epsilon > 0$. Hence the energy density measured by a  static observer is arbitrarily large for  adequately small values of $\epsilon$. Such an observer would  most certainly be burnt upon an attempt to remain at rest at a~very small distance $\epsilon$ above the horizon. In this section we investigate, whether the CBR could influence observers in some more realistic astrophysical context.

% ----------------------------------------------------------
\subsection{$T^{(\mu)(\nu)}$ in a boosted frame}
\label{sub:boost}
Having calculated the static observer's orthonormal tetrad photon density $T^{(t)(t)}$, we may ask about the component $T^{(t')(t')}$ in other local orthonormal frames. In new coordinates corresponding to radial motion, the radiation tensor is expressed as
\be
T^{(\mu')(\nu')} = \Lambda^{(\mu')}_{(\mu)}  \Lambda^{(\nu')}_{(\nu)} T^{(\mu)(\nu)} \ \text{,}
\ee
where $\Lambda$ is a Lorentzian boost matrix in the radial direction, parametrized with velocity $v$. After some algebra one finds that
\be
T^{(t')(t')} = \frac{2 \pi I_\infty \left[(1+\cos a_0)(1 + v^2) + v \sin^2 a_0  \right]}{(1- v^2) g_{tt}^2} \ \text{,}
\label{eq:boosted}
\ee
which becomes
\be\label{eq:tmunuv}
T^{(t')(t')} \approx \frac{27M^2 \pi I_\infty }{r^2 |g_{tt}|} \frac{1+v}{1-v} \approx \frac{1+v}{1-v} T^{(t)(t)}
\ee
close to the horizon. The last formula is particularly significant, since it shows that for any $-1 < v \leq 1$ the singularity of energy density remains and may disappear only for $v = -1$, corresponding to infalling with the speed of light ($c = 1$). However, for a particle in a~geodesic motion, $v = -1$ is exactly the limit of the velocity as the particle approches $r = 2M$.

% -----------------------------------------------------
\subsection{Energy density for  non-static geodesic  observers}
\label{sub:energy}

From Eq. (\ref{eq:tmunuv}) we see that the neccessary condition for the energy density measured by the observer to remain finite at the horizon is that observer's velocity goes to the speed of light at the horizon relative to static observers. Considering a~freely-falling observer following a~geodesic trajectory, we know this to be true. We now check whether the free fall assumption is also a sufficient condition for  finite energy density.
\subsubsection{Radial free fall}
The four-velocity of an observer in a radial free-fall motion is
\be
\mathbf{u}_{FF} = [-(1 -2M/r)^{-1},-\sqrt{2M/r},0,0] \ \text{.}
\ee
We evaluate the energy density as
\be
\rho = T^{\mu \nu} u_\mu u_\nu \ \text{.}
\ee
The following formula can be derived:
\be
\rho_{FF}(r) = 2 \pi I_\infty\frac{ 1 + b_0 + \frac{2M}{3r}(1 + b^3_0) - 2 \sqrt{\frac{2M}{r}} (1 - b^2_0)}{(1-2M/r)^3} \ \text{,}
\label{eq:densityFF}
\ee
where $b_0 = \cos a_0$.
In the limit of $r \rightarrow 2M$ this results in a~``[0/0]'' type of indeterminacy, which can be evaluated to give the \textit{finite} value 
\be\label{eq:9.7}
\rho_{FF}(2M) = \left(  \frac{3}{2}\right)^8 \pi I_\infty \ \text{,}
\ee
which is only about 10 times more than the CBR energy density measured at infinity by the static observer. This is of course very small at the present time, but not so small in the early Universe soon after decoupling of matter and radiation.     

\subsubsection{Circular geodesic}
\begin{figure}[htbp!]
\begin{center}
\includegraphics[trim = 0mm 0mm 0mm 0mm, height=5.9cm]{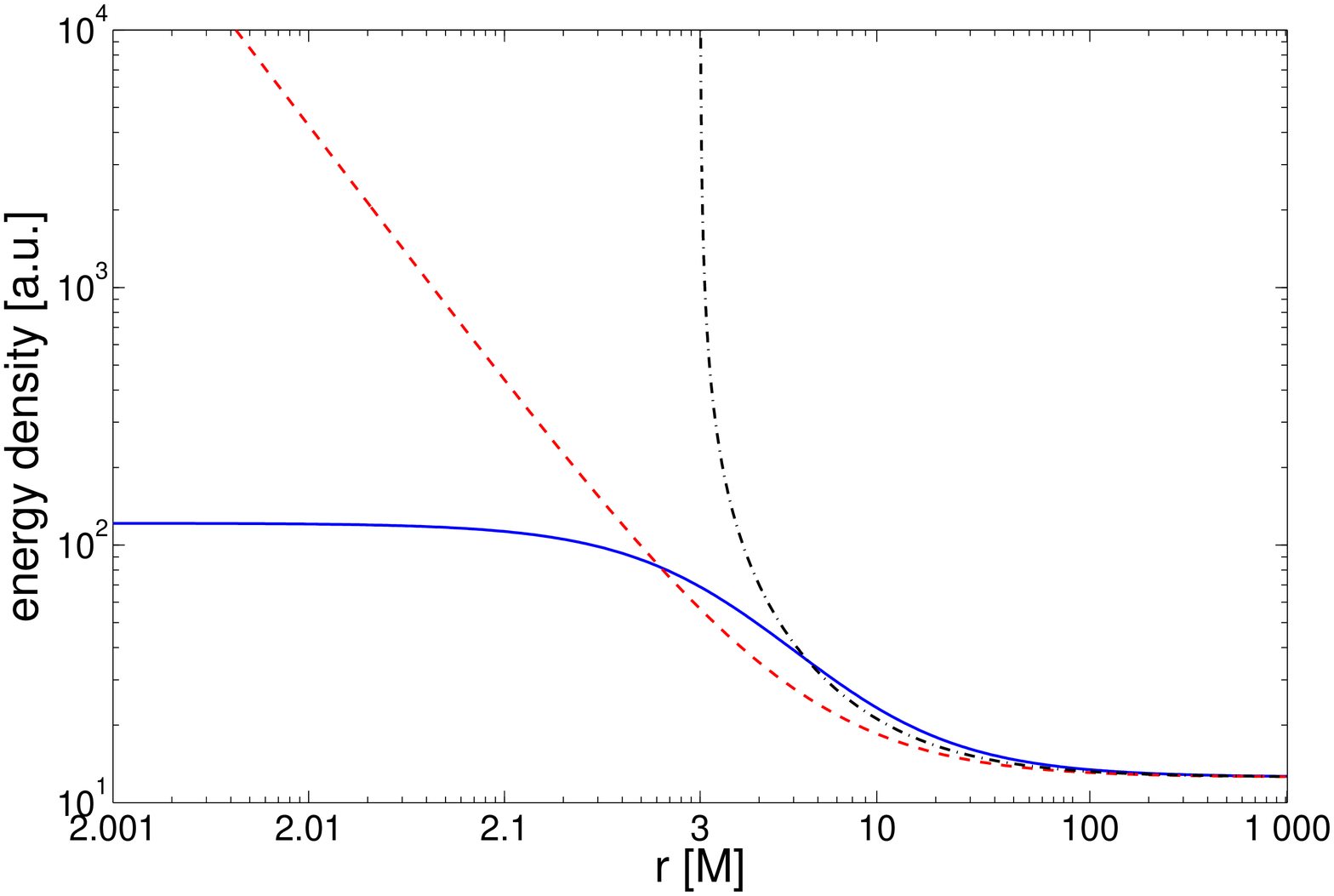}
\caption{Energy density as measured by radially freely-falling observers  (continuous line), static observers  (dashed line) and observers on a  circular geodesic orbit (dash-dot line).}
\label{fig:enDensity}
\end{center}
\end{figure}

Similarly, we can evaluate the energy density as measured by an observer executing a~circular geodesic orbit, i.e. moving with four-velocity
\be\label{eq:circ_velocity}
\mathbf{u}_{CO} = (1 -3M/r)^{-1/2}\left[-1,0,0,M^{1/2}r^{-3/2}\right] \ \text{.}
\ee
The corresponding energy density
\be\label{eq:circ_rho}
\rho_{CO}(r) = \frac{(1-2M/r)T^{(t)(t)} + \frac{M}{r^5}T^{(\phi)(\phi)} }{1-3M/r }
\ee
diverges at $r \rightarrow 3M$. In Fig. \ref{fig:enDensity} the radial dependence of the energy density for static, free-falling and circular geodesic observers are compared. The divergence of the energy density for the circularly moving observers at $r=3M$ is rather dramatic. Does this mean a~spacetime singularity will occur there because of the infalling CBR radiation?

There is an apparent singularity in the four-velocity there, Eq. (\ref{eq:circ_velocity}), and this is the root of the problem.  It occurs because no circular timelike geodesic orbits (stable or unstable) are allowed in Schwarzschild spacetime at or below the photon orbit $r = 3M$.  No massive particle can execute circular geodesic orbit at $r = 3M$ - this demands that the particle moves with the speed of light. Indeed a~circle at  $r = 3M$ is a~null geodesic line, a photon trajectory. So there
  really is no circular timelike geodesic orbit of this kind.  The zero in the denominator of Eq. (\ref{eq:circ_rho}) comes from the four-velocitygiven by  $\mathbf{u}_{CO}$, Eq. (\ref{eq:circ_velocity}), and not from the intrinsic nature of the stress-energy tensor. \\
  
Actually, coordinate singularity is not needed to experience a~divergent energy density of the CBR. Considering 
 the situation far from the black hole, $a_0 = 0$ in Eq. (\ref{eq:boosted}), then for $|v| \rightarrow 1$ we also have
 an infinite value of $T^{(t')(t')}$. Such an observer would indeed experience an unbounded energy density, but this divergence is really because of the extreme velocity and not because of the spacetime curvature. The situation is similar for the circular motion and the $r = 3M$ limit - the infinity is more because of the extreme velocity than because of spacetime curvature effects. So the spacetime is not singular at $r = 3M$, despite what Fig. \ref{fig:enDensity} seems to suggest. We discuss these implications further in Section \ref{sec:discussion}.

% ----------------------------------------------------------
\subsection{Radiation flux and saturation velocity}
\label{sub:force}

In reality, the test particle would not follow a geodesic trajectory exactly. Instead, its motion would be influenced by the radiation field, the energy of which diverges, at least in the static observer's frame. It takes infinite force pointing outwards to prevent the particle from crossing the horizon with the velocity of light, yet this may be the case for the considered radiation field. Being singular at $r = 2M$, it may result in an asymptotically infinite force.\\

\begin{figure}[htbp!]
\begin{center}
\includegraphics[trim = 0mm 0mm 0mm 0mm, height=5.9cm]{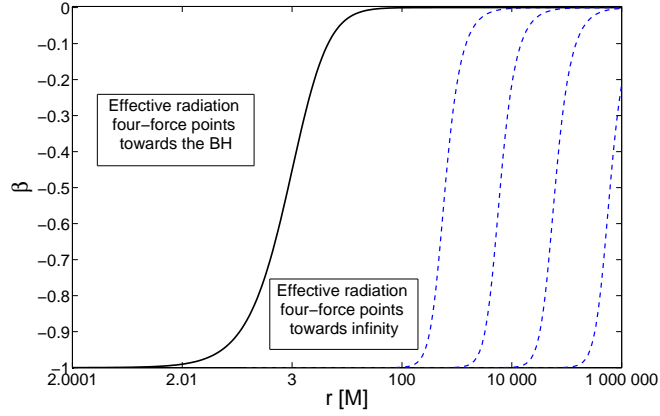}
\caption{The zero flux $\beta_{F0}$ line (continuous) and the saturation velocity $\beta_s$ lines for $M \cdot T^4$, $M = 10^{10} \Msun$, $T = 10^3 T_0$ and  $M \cdot T^4$ smaller by factor of $10^2$, $10^4$, $10^6$.}
\label{fig:saturation}
\end{center}
\end{figure}

Following \citet{abramowicz83}, for purely radial motion, we find the expression for the radial radiation flux
\bea
\label{eq:RadialFlux}
F^r(r ,\beta) &=& h^r_{\ \mu}T^{\mu \nu} u_\nu  \\ \nn
 &=& \frac{(1+\beta^2)T^{(t)(r)} -\beta\left(T^{(t)(t)}+T^{(r)(r)} \right)}{(1-2M/r)^{-1/2}(1-\beta^2)^{3/2}} \ \text{,} \\ \nn
\eea
where $h^\rho_{\ \mu}$ is the projection tensor and the velocity of the particle is parametrized as
\be
\label{eq:beta}
\beta =\pm \left[ \frac{-g_{rr}\left(u^r \right)^2}{g_{tt} \left(u^t\right)^2} \right]^{1/2} \ \text{.}
\ee\\
Equation (\ref{eq:RadialFlux}) can be decomposed into two parts: \textit{radiation pressure}, proportional to $T^{(t)(r)}$, and \textit{radial drag}, proportional to $T^{(t)(t)}+T^{(r)(r)}$. The latter disappears for a  static observer ($\beta = 0$), and always acts against the motion. In the considered context, for an infalling particle radiation pressure acts by accelerating the particle, yet the drag has a~decelerating effect. As discussed by \citep{abramowicz90}, there is a~velocity $\beta_{F0}$ for which radiation pressure balances drag, i.e. effective radiation four-force equals zero:
\be
F^r(r, \beta_{F0}) = 0 \ \text{.}
\ee
The radial dependence of the $\beta_{F0}$ is plotted with a thick continuous line in Fig. \ref{fig:saturation}. If the particle moves towards the black hole faster than $\beta_{F0}$, the effective radiation four-force, dominated by the drag component, acts against the motion. An  interesting quantity is the \textit{saturation velocity} $\beta_s$, for which radial radiation drag is not only strong enough to dominate the radiation pressure term, but also balances the effective gravity. Thus, it is the radial velocity for which the particle does not instantaneously accelerate (strictly, $\dd \beta / \dd r = 0$ is implied for $\beta_s \neq 0$, see \citep{abramowicz90}). Obviously $|\beta_s| \geq |\beta_{F0}|$.  While it can be conceptually counter intuitive, radiation coming from behind will be slowing down the particle for sufficiently large $\beta$, $1 > |\beta| > |\beta_s|$. Exact values of $\beta_s$ can be calculated from the algebraic equation
\be
\frac{m_p \sigma_T r^2}{2cGM}(1 - \beta_s^2)F^r(r ,\beta_s) = 1 \ \text{,}
\ee
where $F^r(r,\beta)$ is given in Eq. (\ref{eq:RadialFlux}), $m_p$ is the test particle (proton) mass, and $\sigma_T$ is Thomson cross section (we only account for the Thomson scattering process). We evaluated the saturation velocity for a black hole of $10^{10}$ solar masses and CMB temperature equal to $10^3 T_0$, $T_{0} = 2.726$ K. In Fig. \ref{fig:saturation} it is shown as the dashed line closest to the $\beta_{F0}$ plot. Other subsequent dashed lines represent the factor $MT^4$ diminished by 2, 4 and 6 orders of magnitude. The particular thing to notice is that at the horizon both  $\beta_{F0}$ and $\beta_s$ go to $\beta = -1$.
If we put the limits of $T^{(\mu) (\nu)}$ at $r \rightarrow 2M$ into Eq. (\ref{eq:RadialFlux}), which we can do if the limit of $F^r$ is unique, we find
\be
F^r \approx - \frac{\pi I_\infty}{[(1-2M/r)(1- \beta^2)]^{1/2}} \frac{1+\beta}{1-\beta} \ \text{.}
\label{eq:RadialFluxH}
\ee
Equation (\ref{eq:RadialFluxH}) indicates, that the radiation flux is either zero (possible for $\beta = -1$) or negative, i.e., directed in the black hole direction - radiation pressure wins over the radiation drag force. This means that the radiation drag influence \textit{cannot} prevent the test particle from crossing the horizon with a~velocity of light, it can only add acceleration in the direction of motion. Hence, close to the horizon, the energy density measured by the observer in free fall \textit{overestimates} the energy density measured by the observer whose trajectory is influenced by the radiation field.

% ----------------------------------------------------------
\subsection{Equations of motion}
Consider the test-particle equation of motion under the radiation four-force $f^\mu$ influence in a Schwarzschild spacetime, \citep[see][]{abramowicz90, stahl12}.
The motion is governed simply by
\be
 f^{\mu} = m_p a^{\mu} = m_p u^\nu \nabla_\nu u^\mu \ \text{,}
\ee
where $u^\mu$ is the particle's four-velocity and $a^\mu$ is the four-acceleration. This can be rewritten as
\be
\frac{1}{r_G}\frac{\dd u^\mu}{ \dd \tau} = \frac{\sigma_T}{m_p c^3} F^\mu -  \Gamma^{\mu}_{\: \: \nu\rho} u^\nu u^\rho \ \text{,}
\label{eom1}
\ee
where $r_G = GM/c^2$, $\Gamma^{\mu}_{\: \: \nu\rho}$ are Christoffel's symbols and $F^\mu$ is the radiation flux and $ \dd \tau = \dd s/ r_G$ is a dimensionless line element. The flux is related to the radiation stress-energy tensor $T^{\nu\rho}$ according to
\be
F^{\mu} = h^\mu_{\:\:\nu} \, T^{\nu\rho}\, u_\rho \  \text{,}
\ee
where $h^\mu_{\:\:\nu}$ is the projection tensor orthogonal to the particle's four-velocity.
We introduce the dimensionless quantity
\be
\widehat{T^{\mu \nu}} = \frac{T^{\mu \nu}}{\pi I(r)} \ \text{,}
\ee
to find
\be
\frac{\dd u^\mu}{\dd \tau} = D \left( \frac{M}{\Msun}\right)  \left( \frac{T_{CBR}}{2.726}\right)^4 \frac{h^\mu_{\:\:\nu} \widehat{T^{\nu\rho}} u_\rho }{ (1-2M/r)^2}  \,  - r_G \Gamma^{\mu}_{\nu\rho} u^\nu u^\rho \text{.}
\label{eq:eom1b}
\ee
$\Msun$ denotes mass of the Sun. The dimensionless quantity $D$ present in the above equation can be evaluated to be equal to
\be
D = \frac{\pi G \Msun \sigma_T \sigma 2.726^4}{m_p c^5} = 2.145 \cdot 10^{-29} \ \text{.}
\ee
While the factor
\be
\left( \frac{M}{\Msun}\right)  \left( \frac{T_{CBR}}{2.726}\right)^4
\ee
may be as large as $10^{20}$ in astrophysically relevant situations. A crude comparison between the two components of the right hand side of Eq. (\ref{eq:eom1b})
suggests little impact of the CBR on the particle dynamics in the black hole vicinity. This was confirmed with numerical calculations of test particles trajectories using the codes described in \citep{stahl12,wielgus12}.

% ------------------------------------------
\subsection{Far away from the black hole}
Fig. \ref{fig:saturation} shows that radiation drag dominates far away from the black hole. At large distance from the black hole, i.e., $M/r \rightarrow 0$ and $a_0 \rightarrow 0$, we find the nonzero $T^{(\mu) (\nu )}$ components
\be
T^{(t)(t)} = 4 \pi I_\infty \ \text{,}
\ee
\be
T^{(r)(r)} = T^{(\theta)(\theta)} = T^{(\phi)(\phi)} = \frac{4}{3} \pi I_\infty  \ \text{.}
\ee
The flux is zero, as the radiation is isotropic. A moving particle experiences a radiation drag force that can be represented as
\be
f^r =  \frac{\sigma_T \beta  \left[ T^{(t)(t)} + T^{(r)(r)} \right]}{c(1-\beta^2)^{3/2}} = \frac{16 \pi \sigma_T I_\infty \beta }{3c(1-\beta^2)^{3/2}}
\ \text{.}
\ee
Without loss of generality we assumed here a radial motion in our spherical coordinates system. Equation (\ref{eq:beta}) simplifies to $\beta = u^r/cu^t$. The drag force $f^r$ always acts against the direction of motion and is proportional to the velocity $\beta$, therefore it constitutes a motion damping effect forcing the particles to remain at rest with respect to the CMB frame. Under the influence of the constant force, a  particle will only accelerate until the saturation velocity $\beta_s$ is reached, for which accelerating force is balanced by the drag force. In cgs units this means that protons moving with velocity $\beta$ experience a drag force of
\be
f^r = 1.164 \frac{\beta}{(1-\beta^2)^{3/2}} \cdot 10^{-36} \left( \frac{T_{CBR}}{2.726}\right)^4 \ \text{[dyn] \ .}
\ee
While this force may be negligible today, shortly after the recombination it was of order of 1 dyn per 1 mole of hydrogen gas and could have significant impact on the gas dynamics.

\subsection{Influence on the infalling observer}

At this point we are able to estimate the amount of energy deposited on the observer plunging into the black hole by the CBR field. We take two strong assumptions to get a~rather crude overestimation. First, we assume free fall motion of the observer (see Subsection \ref{sub:force}). Second, we assume that the observer absorbs all of the measured CBR energy density, which is a good approximation only very close to the horizon, where the  CBR is hotter than the observer. Hence, we simply integrate the energy density along the observer trajectory,
\be
\Delta Q =  \chi \int  T^{\mu \nu} u_\mu u_\nu \dd \tau \ \text{,}
\ee
where constant $\chi$ combines specific heat and surface area of the infalling object. Assuming free fall motion we have
\be
\dd \tau = -\sqrt{\frac{r}{2M}} \dd r
\ee
and
\be
\Delta Q =  \chi \int^{r_0}_{2M}  \rho_{FF}(r) \sqrt{\frac{r}{2M}} \dd r  \approx \frac{\chi \rho_{FF}(2M)(r_0 - 2M)}{(r_0/2M)^{-1/2}} \ \text{.}
\ee
The energy density $\rho_{FF}(r)$ is given by the Eq. (\ref{eq:densityFF}) and plotted in Fig. \ref{fig:enDensity}. We are only interested in the region close to horizon, i.e., $r_0 > 2M$ but $r_0 \approx 2M$. This integral is clearly finite. It is also small -- the infalling observer \textit{will not} be burnt by the CBR energy, unlike the static observer in the horizon vicinity.

% % % % % % % % % % % % % % % % % % % % % % % % % % % % % % % % % % % % % % % % % % % % % % % % %
\section{Back-reaction on spacetime geometry}\label{sec:back}
% % % % % % % % % % % % % % % % % % % % % % % % % % % % % % % % % % % % % % % % % % % % % % % % % % % % % % % % % % %
So far we assumed that the radiation may influence observers, but its energy is sufficiently small that one may neglect its back-reaction effect on the metric. Hence, we were using the Schwarzschild spacetime, being a vacuum solution to the Einstein field equation. However, the  Ricci tensor components do not equal zero exactly, because of the radiation field. The Ricci scalar remains zero, see Eq. (\ref{eqn:trace}). From the Einstein field equations with Ricci scalar set to zero, we have
\begin{equation}\label{EFE}
R_{\mu \nu}  = \kappa T_{\mu \nu} \ \text{.}
\end{equation}
To estimate the Ricci curvature, we  calculate the following scalar $\mathcal{R}$:
\be\label{EFEscalar}
\mathcal{R} = R^{\mu \nu}R_{\mu \nu} = \kappa^2 T^{\mu \nu}T_{\mu \nu} = \kappa^2 \pi^2 I^2(r) W(\cos a_0) \ \text{,}
\ee
where $\kappa = 8 \pi $ and $W(b)$ is a polynomial
\be
W(b) = \frac{2}{3}(b+1)^4(b^2 -4 b + 5) \ \text{.}
\ee
The $W(b)$ polynomial has a zero of multiplicity 4 for $b = -1$, which corresponds to the location of the horizon, $ a_0 = \pi$. Then one finds
\be
\mathcal{R}(r) = \frac{2}{3}\kappa^2 \pi^2 I_\infty^2 (\cos^2 a_0 -4 \cos a_0 + 5) \left[\frac{1+ \cos a_0}{1 - 2M/r} \right]^4
\label{eq:invariantR}
\ee
and the factor in square brackets at the horizon limit is evaluated in the Eq. (\ref{eq:BracketFactor}). Overall we find that $\mathcal{R}$ at the horizon of the Schwarzschild black hole is finite and obeys
\be \label{eq:finite_scalar}
\mathcal{R}(2M) = 10 \left( 3/2\right)^{11} (8 \pi^2 I_\infty)^2  = 4.014 \cdot 10^{-118} \left( \frac{T_{CBR}}{2.726}\right)^8\ \textit{,}
\ee
so it does not depend on the black hole mass.\\

We may evaluate the importance of the radiation to the underlying spacetime geometry by comparing the magnitudes of the Ricci and Weyl tensor components of the Riemann tensor. Because the Schwarzschild solution is a~vacuum solution, the Kretschmann scalar $K
: =R^{abcd}R_{abcd} $ %= C^{abcd}C_{abcd}$
simply measures the magnitude of the Weyl tensor. It has the value
\be
\label{eq:WeylComp}
K(r) = C^{abcd}C_{abcd} = \frac{48M^2}{r^6} \ \text{,}
\ee
thus
\be
K(2M) = 1.576 \cdot 10^{-21} \left( \frac{M}{\Msun}\right)^{-4} \ \text{.}
\ee
The ratio between these two quantities (Eq. \ref{eq:invariantR}, Eq. \ref{eq:WeylComp}) indicates whether Ricci component of the Riemann tensor can be neglected in comparison to the Weyl component and therefore whether utilizing vacuum solution to the Einstein field equations is appropriate
\be
\frac{\mathcal{R}(2M)}{K(2M)} = 2.547 \cdot 10^{-97} \left( \frac{M}{\Msun}\right)^{4}  \left( \frac{T_{CBR}}{2.726}\right)^8 \ \text{.}
\label{eq:RiemannRatio}
\ee
Hence, on the basis of this linear (since $T^{\mu \nu}$ is evaluated assuming Schwarzschild geometry) calculation, under no realistic astrophysical circumstances may the energy of the CBR radiation be large enough to have a dominant influence on the Riemann tensor. On  the contrary, even for supermassive black hole in the Universe filled with CBR as hot as 3000 K, the Weyl tensor dominates by at least 30 orders of magnitude. This justifies neglecting the CBR stress-energy tensor and describing astrophysical black holes by vacuum Einstein field equation solutions. On the other hand,  Eq. (\ref{eq:RiemannRatio}) shows what temperature of the thermal radiation is necessary to invalidate this assumption. This could be relevant for some much more energetic astrophysical processes. Also this does not take into account the non-linearities of general relativity theory; this issues is discussed further in Section \ref{sec:conclude}.

% % % % % % % % % % % % % % % % % % % % % % % % % % % % % % % % % % % % % % % % % % % % % % % % % % %
\subsection{Feeding black hole with CBR}
\label{sec:massdot}
The infalling radiation will increase the mass (or energy, using units where $c =G = 1$) of the central object. We now calculate this effect.
Following Eq. (4.23) from \citep{Accretion1} for the null hypersurface $r_0 = 2M$ we find
\be
\left(\dd M \right)_{rad} = \int \limits_{r_0 = 2M} T^{r}_{\: \: v} \sqrt{-g} \dd \phi \dd \theta \dd v  =  108 \pi^2 I_\infty M^2\dd v \ \text{.}
\label{eq:feeding}
\ee
For a distant static observer, for whom $\dd v = \dd t$, we may cast Eq. (\ref{eq:feeding}) in the form of a~mass increase rate due to the CBR absorption
\be
\frac{\dd M}{\dd t} =  108 \pi^2 I_\infty M^2 \ \text{.}
\label{eq:feeding2}
\ee
This exact result is more than one would get from a simple estimation (ignoring black hole influence on radiation field),
\be
\frac{\dd M}{\dd t} \approx  4 \pi \cdot (2M)^2 I_\infty  = 16 \pi I_\infty M^2
\ee
by a factor of $6.75 \pi \approx 20$. The difference will quantitatively influence black hole evolution models, such as the  one given in \citep{mahulikar12}. We can use Eq. (\ref{eq:feeding2}) to evaluate the mass increase rate in the cgs units
\be\label{eq:grow}
\frac{\dd M}{\dd t} = 1.616 \cdot 10^{-10} \left(\frac{M}{\Msun}\right)^2\left(\frac{T}{2.726}\right)^4 \text{[g/s] \ .}
\ee
In every second supermassive black holes grow thousands of tons just by absorbing the CBR. Nevertheless, since the Eddington accretion rate is of order of $10^{17} M/\Msun$ [g/s], quantity given by Eq. (\ref{eq:grow}) is vanishingly small.

%-------------------------------
\subsection{The Vaidya solution with CBR}
When the evolution of the black hole on cosmological timescales is considered, the non-stationary character of the metric indicated by Eq. (\ref{eq:grow}) must be taken into account. One can represent the effect of the infalling radiation on the spacetime metric by an approximation based on the Vaidya solution to the Einstein field equation \citep{vaidya}. Although the Vaidya solution only accounts for the photon radial motion, neglecting their angular momentum (hence it ignores $T^{\theta \theta}$ and $T^{\phi \phi}$ components), it takes the back-reaction of such an approximated radiation field into account exactly.\\

The line element of the Vaidya ingoing radiation metric is given by 
\be
\dd s^2 = -\left[1 - \frac{2M(v)}{r} \right] \dd v^2 + 2 \dd v \dd r + r^2(\dd \theta^2 + \sin^2 \theta \dd \phi^2) \ \text{,}
\label{eq:vaidya}
\ee
where $v$ is an advanced null coordinate and $M(v)$ is an arbitrary function. Obviously it is increasing with time due to the CBR infall. When $M(v) = M_0 = \text{const}$, Eq. (\ref{eq:vaidya}) reduces to the Schwarzscild spacetime line element in the Eddington-Finkelstein coordinates. Now, if we calculate the $T_{vv}$ component of the radiation stress-energy tensor in Schwarzschild spacetime, expanding the result around $r = 2M$, we find
\be
T_{vv} \approx 27 \pi I_\infty(M/r)^2 \ \text{.}
\ee
For the Vaidya metric to take  proper account of the  presence of the radiation tensor  component $T_{vv}$, one needs to fulfill
\be
T_{vv} = \frac{1}{4 \pi r^2} \frac{\dd M(v)}{\dd v} \ \text{,}
\ee
hence
\be
\frac{\dd M(v)}{\dd v} = 108 \pi^2 I_\infty M^2  \ \text{,}
\label{eq:vaidyaM}
\ee
which is consistent with the finding from Eq. (\ref{eq:feeding}). When provided with a~model of CBR cooling $T(v)$, Eq. (\ref{eq:vaidyaM}) can easily be  integrated to give the equation for the evolution of the black hole mass. The Hawking radiation, \citep[see][]{hawkrad, hawkrad1} has been neglected, since in any known astrophysical context (except for the distant future of the Universe), its influence on the black hole mass evolution is many orders of magnitude smaller than the CBR influence.

% % % % % % % % % % % % % % % % % % % % % % % % % % % % % % % % % % % %
\section{Discussion: a space-time singularity?}\label{sec:discussion}

So is there a spacetime singularity? This is a surprisingly tricky question. Arguably a singularity has the potential to arise when non-linear backreaction effects are taken into account, but this may not occur in astrophysical reality.\\
% but of a rather special type.\\

The Einstein equation (\ref{EFE}) shows that in the static observer's orthonormal tetrad frame,
\begin{equation}\label{eq:rmn}
R^{(\mu) (\nu)}  = \kappa T^{(\mu) (\nu)}.
\end{equation}
Some of the right hand terms diverge at $r = 2M$, by Eqs. (\ref{eq:Ttt})-(\ref{eq:Ttr}), so the left hand sides diverge also. This strongly suggest that a space-time
singularity occurs in this frame when feedback effects are taken into account. This is also true in any other orthonormal frame where $|v| \neq 1$ in the limit $r \rightarrow 2M$ (see Eq.
(\ref{eq:tmunuv})). However it is, in terms of the classification of singularities \citep{sing}, a~non-scalar singularity
because the trace $R$ vanishes (see Eq. \ref{eqn:trace}) and the scalar $R^{ab} R_{ab}$ is finite, see Eq. (\ref{eq:finite_scalar}). That is why the curvature can be finite in the radially freely falling frame, as shown above.\\

In fact it is an intermediate singularity \citep{sing}, that is one that is finite in some orthonormal frames but not in others. This is similar to what happens in some tilted homogeneous cosmologies, \citep{sing1,sing2}, except that here the
situation is reversed: in those cases the Ricci tensor diverges in a~parallel transported orthonormal frame but is finite in a~group invariant frame; here the situation is the other way round. The root of this issue is the divergence of parallel propagated
vectors as opposed to group invariant vectors that always occurs when there is a bifurcate Killing horizon \citep{boyer}, which
of course occurs in the Schwarzschild solution.
Now it is true that if the density $\rho$ has a finite value in one frame $u^a$, one can always make it appear to diverge by a Lorentz transformation from the frame $u^a$ to a~frame $u'^a$ with velocity $v$ that diverges to the speed of light ($|v|/c \rightarrow 1$). The question is whether this is just a~coordinate singularity, or  should be regarded as a physical singularity. That depends on whether the frame $u'^a$ can be regarded as physically meaningful or not. If it is for example the timelike eigenvector  of the Ricci tensor, this would be a spacetime singularity; however that is not the case here. The static observer's frame is geometrically preferred because it is defined by the timelike symmetry; but as explained above, no object can move on those non-geodesic orbits when close enough to $R=2M$,  because that would take unbounded rocket engine power (because the acceleration of the static orbits diverges in this limit). This cannot happen for real physical objects, so in the end this is like  a~coordinate rather than physical singularity. \\ 

However this argument does not apply to tangentially moving particles in circular orbits. The key feature here is shown in Figure 3: the energy density diverges for particles on circular orbits at fixed radius $r$ as $r\rightarrow 3M$. This is not directly due to the event horizon and coordinate singularity  at $r=2M$, as this occurs further out. There exist timelike circular geodesic orbits for $3M + \epsilon$ for all $\epsilon > 0$; they are geodesically complete orbits that represent possible particle motion, although they are unstable.
If we consider releasing a particle at an inward angle $\alpha$ on such an orbit at $r = 3M + \epsilon_0$, for $\alpha=0$ one will have a~circular orbit; for $\alpha = \pi/2$, radial infall; by the existence and uniqueness theorems for geodesics, for small enough $\alpha$, there will be a geodesic path that travels in as slowly as one cares through all values $r = \epsilon$, $0< \epsilon < \epsilon_0$, and the effective radiation energy density on that path will diverge as $\epsilon \rightarrow 0$. This will cause major heating that would indeed be experienced as a~real firewall, and so will lead to release of $\gamma$-radiation. It will also  cause refocussing of timelike geodesics, so if one released a cloud of particles on such orbits, their density would increase and diverge as a conjugate point is reached at a finite affine parameter distance; this non-linear feedback effect has the potential to create a scalar singularity where the energy density diverges and the particle paths end \citep{HE}. This then qualifies as a~physical spacetime singularity.
However, such a~particle motion is not what occurs in realistic accretion situations: although this can occur in principle, the cosmic censor may prevent it happening in practice. What particles in real accretion disks do is that they spiral in until the last stable circular orbit at $r = 6M$ and then they fall in almost radially, when nothing untoward occurs. They do not accumulate at $r = 3M$, and so do not experience this diverging energy density. The potential singularity due to the infalling CBR radiation is probably not realised in real astrophysical contexts. The photon orbit divergence is related to the fact that such a hypothetical observer would have to travel with the speed of light to remain on the circular orbit at r = 3M so the photons that travel
 towards him are very much blueshifted. 
 
 Thus the very reason for the behaviour noted here is the infinite blueshift of infalling photons at the event horizon in the ZAMO frame, accentuated for Killing vector frames with a tangential velocity component. We have calculated the effect for the CBR, but it occurs for any infalling photons --- originating in accretion disk around the black hole, incoming from the host galaxy and other galaxies. In these cases the effect may be non-negligible, see, e.g., Eq. (\ref{eq:9.7}).\\

It is clear that there are indeed potential classical firewall effects that occur for a black hole imbedded in such a~radiation field. They may however be avoided in practice. Susskind et al \cite{Suss93} proposed a principle of Black Hole Complementarity for black holes when quantum effects are taken into account: information is both reflected at the event horizon and passes through the event horizon and can't escape, but no observer can experience both views. According to an external observer,  infalling information heats up a region just above the event horizon which then reradiates it as Hawking radiation, so information is reflected there, while according to an infalling observer, nothing special happens at the event horizon itself, so the information is lost in the interior and no heating takes place near the event horizon. We see now that in the classical case there is also a somewhat similar kind of complementarity, but a bit different: according to an external observer who remains outside,  infalling radiation heats up a region just above the event horizon to the extent that the stress tensor of matter diverges in the rest frame of such observers; they would expect backreaction from this matter tensor to cause a spacetime singularity there. However according to an infalling observer, nothing special happens at the event horizon itself, so there is no reason to believe a singularity will occur there when backreaction is take into account, despite the experiences of the ZAMO observers. Then the key point is that in astrophysically realistic cases, there will be no physical entities corresponding to ZAMO observers: their potential experience is not realised by any family of particles hovering at close to the event horizon for extended periods. The crucial difference seems to be that according to Susskind et al, quantum effects will indeed occur for reference frames that correspond to the physically meaningful existence of ZAMO observers.

% % % % % % % % % % % % % % % % % % % % % % % % % % % % % % % % % % % % % % % % % % % % % % % % % % % % % %

\section{ Conclusions}\label{sec:conclude}
We have examined consequences of the infinite blueshift of the Cosmic Background Radiation photons, measured in the
ZAMO (static) frame at the horizon of a~Schwarzschild black hole. With analytic calculations we investigated several aspects of the CBR interaction with black hole. We found that such effects have typically little importance for the astrophysical processes (even in the past hot CBR era). In particular, provided backreaction effects are not large, these photons do not form a~physically important ``firewall''. %Further work 
However work still needs to be done to check that non-linear backreaction effects associated with circular orbits near $r=3M$, as discussed in the last section, do not win the day. If some mechanism injected matter or photons into such orbits, the effects discussed here would not be negligible.  

While we focused on the CBR field influence, the results can be easily used to evaluate the impact of other symmetric radiation fields, e.g., averaged radiation flux from the stars forming a~spherical galaxy with a~supermassive black hole in its center.
\newpage
\begin{acknowledgements}
We acknowledge support from the Polish NCN grant UMO-2011/01/B/ST9/05439 and the NRF (South Africa). Work of M.W. was partially supported by the European Union in the framework of European Social Fund through the Warsaw University of Technology Development Programme. M.W. also thanks La Lechuza Hostel in Rosario (Argentina) crew and guests for inspiring discussions on the subject of this paper.
\end{acknowledgements}

\end{document}